\documentclass{article}
\usepackage{amsmath,amssymb,amsfonts}
\usepackage{algorithmic}
\usepackage{graphicx}
\usepackage{textcomp}
\usepackage{xcolor}
\usepackage{hyperref}
\usepackage{physics}
\usepackage[]{algorithm2e}
\usepackage{mdframed}
\usepackage{caption}
\usepackage{subcaption}
\usepackage{flafter}
\usepackage{float}
\usepackage{physics}
\usepackage{longfbox}
\usepackage{tabularx}
\makeatletter
\newdimen\@tempdimd
\makeatother
\usepackage{natbib}
\usepackage[T1]{fontenc}
\fboxset{rounded, border-width=0.2pt, padding={0.2ex,0.4ex}}
\usepackage{tikz}
\usetikzlibrary{decorations.pathreplacing}
\usepackage{authblk}

\title{Heuristics for Detecting CoinJoin Transactions on the Bitcoin Blockchain}

\author[1,2]{Hugo Schnoering\thanks{hschnoering@coinshares.com}}
\author[2]{Michalis Vazirgiannis\thanks{mvazirg@lix.polytechnique.fr}}

\affil[1]{\small Coinshares}
\affil[2]{\small Ecole Polytechnique}

\begin{document}

\maketitle

\begin{abstract}
This research delves into the intricacies of Bitcoin, a decentralized peer-to-peer network, and its associated blockchain, which records all transactions since its inception. While this ensures integrity and transparency, the transparent nature of Bitcoin potentially compromises users’ privacy rights. To address this concern, users have adopted CoinJoin, a method that amalgamates multiple transaction intents into a single, larger transaction to bolster transactional privacy. This process complicates individual transaction tracing and disrupts many established blockchain analysis heuristics. Despite its significance, limited research has been conducted on identifying CoinJoin transactions. Particularly noteworthy are varied CoinJoin implementations such as JoinMarket, Wasabi, and Whirlpool, each presenting distinct challenges due to their unique transaction structures. This study delves deeply into the open-source implementations of these protocols, aiming to develop refined heuristics for identifying their transactions on the blockchain. Our exhaustive analysis covers transactions up to block 760,000, offering a comprehensive insight into CoinJoin transactions and their implications for Bitcoin blockchain analysis.

\end{abstract}

\section*{Introduction}

Bitcoin (\citet{nakamoto2008bitcoin}) is a decentralized peer-to-peer network designed to transfer value in the form of bitcoins between participants. These value transfers are encoded in transactions, and the entire sequence of transactions since the network's inception is chronicled on a public and decentralized ledger known as the Bitcoin blockchain. This blockchain is composed of a series of blocks, each containing a set of transactions. For ease of reference, we will often use the block index in the blockchain as a temporal measure. All transactions are visible on the blockchain and can thus be scrutinized. Notably, it is possible to track users' funds, jeopardizing their right to privacy. CoinJoin has emerged as a commonly used method to bolster privacy for Bitcoin users. It involves a collaboration wherein individuals combine their transaction intents into one large transaction. Consequently, tracking individual transfers becomes intricate. Furthermore, these transactions create numerous ties on the blockchain between participants who, ultimately, have no real connection other than having jointly participated in a CoinJoin. As a result, many existing analysis heuristics are confounded. Recognizing these transactions becomes useful when one undertakes the analysis of the Bitcoin blockchain.  A primary challenge in detecting CoinJoin transactions lies in the absence of a dataset of verified CoinJoin transactions. Among the implementations of CoinJoin that have been somewhat successful, we can mention JoinMarket, Wasabi, or Whirlpool. Each protocol has its own specificities, therefore the transactions generated by the different implementations can have very different forms. Moreover, the Wasabi software has undergone major improvements which have significantly changed the shape of the resulting transactions. We will distinguish three versions of the Wasabi software: 1.0, 1.1, and 2.0, which correspond to the different tags in the GitHub repository\footnote{\url{https://github.com/zkSNACKs/WalletWasabi}}.

\paragraph{Related Works} This research topic has been sparsely explored, but we can reference the work of \citeauthor{nopara73} who devised heuristics to detect transactions generated by the softwares JoinMarket, Wasabi (version 1.1) and Whirlpool. The author then used these heuristics to calculate metrics that allow for comparing the different implementations in terms of adoption and efficiency. \citeauthor{stockinger2021pinpointing} also studied the Wasabi (version 1.1) and Whirlpool implementations. The author uses heuristics that are inspired by, but improved upon, Fiscor's heuristics, in order to improve the precision of detection. The author uses the heuristics to detect CoinJoin transactions up to block 689255, primarily to measure the volume of bitcoins that have passed through this type of transactions. The detection of CoinJoin transactions can be useful in future work. For example, \citeauthor{wahrstatter2023improving} uses the distance of a user to a CoinJoin transaction as a feature to detect users involved in a crime. To detect CoinJoin transactions, the author used the heuristics of \citeauthor{nopara73}  and \citeauthor{stockinger2021pinpointing} mentioned previously. In this study, it was demonstrated that this feature improved detection results. Indeed, criminals often use these softwares to cover their tracks and launder bitcoins obtained illegally.

\paragraph{Contributions} The implementations of the different softwares studied are open-source, which allows us to meticulously study the process of creating transactions and to understand the subtleties of each protocol. From this study, we develop heuristics for the following softwares: JoinMarket, Wasabi (versions 1.0, 1.1, 2.0) and Whirlpool. Regarding version 1.1 of the Wasabi software, we developed a heuristic that is significantly different from the work of \citeauthor{nopara73} and \citeauthor{stockinger2021pinpointing}, more faithful to the software's implementation. As for Whirlpool, our heuristic is very similar to those of the previously mentioned studies. However, we also developed a heuristic to detect Whirlpool's Tx0 transactions, which allow a user to participate in a Whirlpool CoinJoin. Finally, we read all the transactions on the chain up to block number 760,000, in order to detect transactions generated by the various softwares. The subsequent sections of this study are structured as follows: in section \ref{sec:first_def}, we introduce general concepts related to Bitcoin and CoinJoin transactions. In section \ref{sec:coinjoin}, we will introduce the softwares we will examine, and we will formulate our heuristics to detect transactions generated by these softwares.  Finally, in section \ref{sec:results}, we showcase the results derived from applying our heuristics to all transactions up to block of index 760,000.

\section{CoinJoin Transactions on Bitcoin}

\label{sec:first_def}

Bitcoin relies on the principles of public-key cryptography. Within this framework,  a user  $u$ can possess or control a set of one or multiple private keys, represented collectively as $\mathcal{K}_u$. Each private key $k \in \mathcal{K}_u$ secures a portion of $u$'s wealth. $u$ must therefore keep its private keys secret, as their knowledge allows to access and control the associated funds. Rather than exposing the private keys to the network, Bitcoin employs a range of derived identifiers, such as public keys, public key hashes, and various other quantities, all obtained from private keys through one-way cryptographic functions or hash functions. These pseudonymous representations of $u$, collectively referred to as \textit{addresses}, can thus be safely shared with others, enabling the other participants to identify $u$.

\paragraph{Transaction Output} A \textit{transaction output} (TXO) $\tau$  is defined by two components: a value $v \in \mathbb{N}$ in satoshis (1 satoshi = $10^{-8}$ bitcoin) and $\ket{p}$ a part of a computer program written in Bitcoin script. $\ket{p}$ is called a \textit{locking script} because  it specifies the conditions under which $\tau$,  and by consequent the associated value $v$, can be spent. 
In most cases, $\ket{p}$ specifies one or several identifiers, derived from some private keys $k_1, ..., k_i$, such that only users that knows $k_1, ..., k_i$ can "unlock" $\ket{p}$ and then spend the output $\tau$. The group of users $u_1, ..., u_j$ that have the knowledge of these keys are said to be the owners of $\tau$. 

\begin{figure}[H]
\centering

\resizebox{0.3\textwidth}{!}{
\begin{tikzpicture}


\node[circle, draw, font=\tiny] (nodein1) at (-2,2) {$\ket{p}_1$};
\node[circle, draw, font=\tiny] (nodein2) at (-2,1) {$\ket{p}_2$};
\node[circle, font=\tiny] (nodeinj) at (-2,-0.5) {$\vdots$};
\node[circle, draw, font=\tiny] (nodeink) at (-2,-2) {$\ket{p}_k$};

\node[circle, draw, font=\Large, minimum size=0.95cm] (nodedelta) at (0,0) {};
\node[circle, draw, font=\Large, minimum size=0.8cm] (nodedelta1) at (0,0) {$\Delta$};

\node[circle, draw, font=\tiny] (nodeout1) at (2,2) {$\ket{p}_1^\prime$};
\node[circle, draw, font=\tiny] (nodeout2) at (2,1) {$\ket{p}_2^\prime$};
\node[circle, font=\tiny] (nodeoutj) at (2,-0.5) {$\vdots$};
\node[circle, draw, font=\tiny] (nodeoutk) at (2,-2) {$\ket{p}_l^\prime$};


\draw[->] (nodein1) -- (nodedelta) node[midway, above,font=\tiny] {$v_1$};
\draw[->] (nodein2) -- (nodedelta) node[midway, above,font=\tiny] {$v_2$};
\draw[->] (nodeink) -- (nodedelta) node[midway, above,font=\tiny] {$v_k$};

\draw[->] (nodedelta) -- (nodeout1) node[midway, above,font=\tiny] {$v_1^\prime$};
\draw[->] (nodedelta) -- (nodeout2) node[midway, above,font=\tiny] {$v_2^\prime$};
\draw[->] (nodedelta) -- (nodeoutk) node[midway, above,font=\tiny] {$v_l^\prime$};

\end{tikzpicture}
}
\caption{Schematic of a transaction $\Delta = (\Delta_\text{in}, \Delta_\text{out})$ with $\Delta_\text{in} = \{ (\ket{p}_i, v_i) \}_{1 \leq i \leq k}$ and $\Delta_\text{out} = \{ (\ket{p}_i^\prime, v_i^\prime) \}_{1 \leq i \leq l}$. Nodes with a single border symbolize TXOs.}
\label{fig:scheme_transaction}

\end{figure}
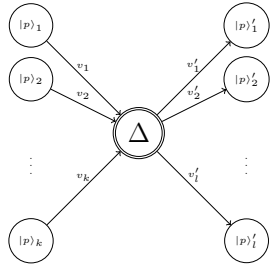

\paragraph{Transaction} A transaction $\Delta$ is characterized by two finite sets of TXOs: the input TXOs, $\Delta_\text{in}$, and the output TXOs, $\Delta_\text{out}$. $\Delta$ symbolizes the transfer of value from the owners of the input TXOs to the owners of the output TXOs. The input TXOs in $\Delta_\text{in}$ provide the required funds for the transaction.  The total input value, denoted as $v_\text{in}(\Delta)$, is defined as the sum of the values $v$ for all TXOs $(v, \ket{p})$ present in $\Delta_\text{in}$, i.e. $\sum_{(v, \ket{p}) \in \Delta_\text{in}} v$.  Subsequently, $v_\text{in}(\Delta)$ is divided and allocated among the TXOs that form $\Delta_\text{out}$ with a total value  $v_\text{out}(\Delta) \triangleq \sum_{(v, \ket{p}) \in \Delta_\text{out}} v $. The total input value has to exceed or equal the total output value\footnote{We do not consider Coinbase transactions in this work.}. In other words, users can not spend more value than they have agreed to include in the transaction $\Delta$. Mathematically, this condition can be expressed as: $v_\text{in}(\Delta) \geq v_\text{out}(\Delta)$. The difference $v_\text{in}(\Delta) - v_\text{out}(\Delta) \geq 0$ represents the network fee paid to the user, the miner, responsible for incorporating the transaction into the blockchain. Given that the number of transactions per unit of time is bounded by the protocol's characteristics, miners prioritize transactions with the highest network fees. Thus, the network fee serves as a bidding tool to secure a position within the chain more rapidly. These fees can vary significantly depending on network congestion, i.e., based on the number of users aiming to execute a transaction simultaneously. We can see in Figure \ref{fig:scheme_transaction} a schematic representation of a transaction. It is not necessary for the inputs to belong to the same user, the same goes for the outputs.

\begin{figure}[H]
\centering
\resizebox{0.3\textwidth}{!}{
\begin{tikzpicture}

\node[circle, draw, font=\tiny, fill=gray!20] (nodein1) at (-2,2) {$\tau_1$};
\node[circle, draw, font=\tiny, fill=gray!20] (nodein2) at (-2,1) {$\tau_2$};
\node[circle, font=\tiny] (nodeinj) at (-2,-0.5) {$\vdots$};
\node[circle, draw, font=\tiny, fill=gray!20] (nodeink) at (-2,-2) {$\tau_k$};

\node[circle, draw, font=\Large, minimum size=0.95cm] (nodedelta) at (0,0) {};
\node[circle, draw, font=\Large, minimum size=0.8cm] (nodedelta1) at (0,0) {$\Delta$};

\node[circle, draw, font=\tiny, fill=white!20] (nodeout1) at (2,1) {$\ket{p}^\prime$};
\node[circle, draw, font=\tiny, fill=gray!20] (nodeout2) at (2,-1) {$\ket{p}$};


\draw[->] (nodein1) -- (nodedelta);
\draw[->] (nodein2) -- (nodedelta);
\draw[->] (nodeink) -- (nodedelta);

\draw[->] (nodedelta) -- (nodeout1) node[midway, above,font=\tiny] {$v^\prime$};
\draw[->] (nodedelta) -- (nodeout2) node[midway, above,font=\tiny] {$v$};

\end{tikzpicture}
}
\caption{Schematic of a payment transaction $\Delta$ with change. Gray scripts / TXOs belong to user $u$, and the white script belongs to $u^\prime$.}
\label{fig:scheme_payment_transaction_change}

\end{figure}
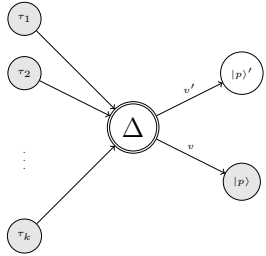

\paragraph{Payment Transaction} Consider two users, $u$ and $u^\prime$, where $u$ intends to transfer a total value $v$ to $u^\prime$. Let us assume that $u$ has control over a set of TXOs, denoted as $\tau_1, \tau_2, ..., \tau_k$ with a cumulative value surpassing $v$ plus the network fee $\theta$. To effectuate the transfer of value $v$ to $u^\prime$, $u$ can construct a transaction $\Delta = (\Delta_\text{in}, \Delta_\text{out})$ where $\Delta_\text{in}$ comprises the set $\{\tau_1, \tau_2, ..., \tau_k \}$, and

\begin{equation}
\Delta^{out} = \left\{
\begin{array}{cc}
( \underbrace{(v, \ket{p}^\prime )}_\text{payment output} ) & \text{if } v_\text{in}(\Delta) = v + \theta, \\
\\
((v, \ket{p}^\prime), \underbrace{(v_\text{in}(\Delta) - v - \theta, \ket{p}}_\text{change output}) & \text{else.} \\
\end{array}
\right.
\end{equation}

where the locking scripts $\ket{p}$ and $\ket{p}^\prime$ are controlled by $u$ and $u^\prime$, respectively. In the first case, the total input value precisely matches the amount (plus the network fees) that $u$ intends to transfer to $u^\prime$. However, in the second case, the total input value exceeds $v + \theta$. In such instances, $u$ generates an additional output to account for the change from the payment. This transaction is depicted in Figure \ref{fig:scheme_payment_transaction_change}.

\paragraph{CoinJoin Transaction} All transactions are visible on the blockchain and can thus be investigated. In the previous example, even though it might be challenging for an observer to determine which output is the payment and which output is the change, especially if $u$ and $u^\prime$ have created two new scripts / identities specifically to receive the payment and change. However, it remains relatively straightforward to track how $u$ spends its bitcoins. The concept of a CoinJoin transaction was first described by Gregory Maxwell in 2013 on the BitcoinTalk forum. Imagine we have a set of users $(u_1, u_2, ..., u_n)$ with transaction intentions $\Delta_1, \Delta_2, ..., \Delta_i$ such that each user $u_i$ wishes to send a value $v_{i}^\prime$ to a user $u_i^\prime$ and receive a change $v_i$, i.e.

\begin{equation}
\small
 \Delta_{i, \text{out}} =   \left( (v_i, \ket{p}_i), (v_{i}^\prime, \ket{p}_{i}^\prime) \right)  
\end{equation}

If each user decides to make their transaction independently, then an observer can determine how each $u_i$ spends their funds, and from whom each $u_{i}^\prime$ receives funds. Privacy can be enhanced by merging the various intentions into a large transaction $\Delta$\footnote{For simplicity's sake, we will not discuss the order of the TXOs in the sequences $\Delta_\text{in}$ and $\Delta_\text{out}$, but it should be imagined that in our case a non-trivial permutation is applied to mix the outputs, and the same for the inputs.},

\begin{align}
\small
& \Delta_\text{in} = \underset{1 \leq i \leq n}{\oplus} \Delta_{i, \text{in}} \\
& \Delta_\text{out} = \underset{1 \leq i \leq n}{\oplus} \Delta_{i, \text{out}}
\end{align}

where $\oplus$ denotes the concatenation of finite-length sequences. Since the parties have combined their inputs and outputs into a single transaction, we refer to this as a CoinJoin transaction. In this manner, to analyze the different payments, the first task will be to determine which inputs produced which outputs, i.e., to find the original $(\Delta_{i, \text{in}} , \Delta_{i, \text{out}})$, which can be complex. Moreover, suppose now that $\forall i, \ v_{i}^\prime = v$ and $\forall i, \ v_\text{in}(\Delta_i) \geq v$. The payments will be indistinguishable, making it impossible for an observer to ascertain with certainty, without additional information, which $\Delta_{i, \text{in}}$ funded which output of value $v$\footnote{However, this might not necessarily be true for change outputs, which will often have a different value; the challenge of linking inputs to change is sometimes referred to as the sudoku problem.}

\section{Detection Heuristics}

\label{sec:coinjoin}

There exist various other forms of CoinJoin. For instance, not all participants are required to use the same value $v$ (Wasabi 2.0) as long as, for each $v$, there remains ambiguity about the origin of the funds for outputs of value $v$. In other variations, each $u_{i}^\prime$ might receive multiple TXOs of the same value or of different values (Wasabi 1.1 / 2.0). In such cases, for maximum privacy gain, these TXOs should be protected by distinct scripts derived from different private keys. The indistinguishable output TXOs will henceforth be termed as post-mix, and their values will be referred to as post-mix denominations. CoinJoin transaction arises from the collaboration of multiple participants, and the construction of $\Delta$ can either be coordinated by one of the participants (JoinMarket) or by a third-party coordinator (Wasabi / Whirlpool). We will refer to the process of creating $\Delta$ as a round. Lastly, it should be noted that the primary use-case is not to obscure a payment but to break the traceability of one's TXOs, i.e., in most cases, $u_{i}^\prime$ is another identity of the participant $u_i$. CoinJoin transactions also challenge the common-input-ownership heuristic (\cite{meiklejohn2013fistful}) which posits that for a transaction $\Delta$, all scripts involved in $\Delta_\text{in}$ belong to the same user. This heuristic is frequently employed in Bitcoin blockchain research/analysis but leads to false conclusions when applied to a CoinJoin transaction.\\ 

In the following, we will use the following two notations extensively: the number of input scripts $n_\text{scripts, in} \triangleq | \{ \ket{p} \}_{(\ket{p}, v) \in \Delta_\text{in}} |$ and the number of output scripts $n_\text{scripts, out} \triangleq | \{ \ket{p} \}_{(\ket{p}, v) \in \Delta_\text{out}} |$.

\subsection{JoinMarket}

JoinMarket\footnote{\url{https://github.com/JoinMarket-Org/joinmarket-clientserver} version 0.3} is a software that enables users, termed as takers, to coordinate CoinJoin rounds. The construction of a round is predicated on the existence of an order book of liquidity made available to takers by other passive participants, known as makers. The orders in the order book are posted by the makers. Each order defines a range $[d_\text{min}, d_\text{max}]$ of CoinJoin denominations which the maker is willing to participate in, a fee that the taker will have to pay to mobilize the maker's liquidity, and the contribution to the network fee that he is willing to cover in a transaction. When a taker wishes to initiate a round of denomination $d$, he will select a number of offers from makers in the order book that are likely to accept participating in his round. He contacts the selected makers who can either accept or decline his offer. A maker agrees to the offer by showing the TXOs he intends to use to fund his participation, a script to receive the post-mix of denomination $d$, and a second script to receive the change. If enough makers have responded, the taker takes responsibility for constructing the transaction by including, as inputs, his TXOs and the TXOs of all the makers who have agreed. In output, each maker receives a post-mix output of denomination $d$ and a change. The taker gets a post-mix output and a potential change. From the resulting transaction $\Delta$, we can infer a few properties. We will respectively denote $n$ and $d$ as the number of participants and the post-denomination.

\begin{enumerate}
\small

\item  Each participant receives exactly one post-mix output and a potential change if it isn't negligible. Therefore, we can easily bound the number of outputs.
\begin{equation}
\small
| \Delta_\text{out} | \leq \underbrace{n}_{\substack{\text{Number of post-mix} \\ \text{outputs}}} + \underbrace{n}_{\substack{\text{Maximum number of} \\ \text{change outputs}}}
\end{equation}
or equivalently,
\begin{equation}
\small
\boxed{
n \geq \frac{| \Delta_\text{out} |}{2}
}
\label{eq:join_upper_bound_outputs}
\end{equation}

\item From the previous point, we also deduce that
\begin{equation}
\small
\boxed{
 n = \underset{v^\prime \in \mathbb{N}}{\text{max}}\sum_{(\ket{p}, v) \in \Delta_\text{out}} 1_{v = v^\prime}
 }
 \label{eq:join_estimation_n}
\end{equation}
and
\begin{equation}
\small
d \in \left\{ v^\prime \in \mathbb{N} \ | \ \sum_{(\ket{p}, v) \in \Delta_\text{out}} 1_{v = v^\prime} = n \right\}
\end{equation}

\item Since there are $n$ participants, the input TXOs are protected by at least $n$ distinct scripts. Moreover, the protocol requires that there are at least 3 participants. These two properties are written as:
\begin{equation}
\small
\boxed{
3 \leq n \leq n_\text{scripts, in}
}
\label{eq:join_lower_bound_input_scripts}
\end{equation}

\item To fulfill its function, all outputs must be protected by distinct scripts:
\begin{equation}
\small
\boxed{
| \Delta_\text{out} | = n_\text{scripts, out}
}
\label{eq:join_diff_scripts}
\end{equation}
\end{enumerate}

\vspace{0.5cm}

Given a transaction $\Delta^\prime$, we will assume that $\Delta^\prime$ is a JoinMarket type CoinJoin transaction if conditions \ref{eq:join_upper_bound_outputs}, \ref{eq:join_lower_bound_input_scripts}, and \ref{eq:join_diff_scripts} are met. It's important to note that $n$ is not known a priori; we will therefore estimate it using equation \ref{eq:join_estimation_n}.

\label{sec:joinmarket}

\subsection{Wasabi 1.0}

\label{sec:wasabi1}

Wasabi 1.0 is a software implemented in version 1.0 of Wasabi Wallet\footnote{\url{https://github.com/zkSNACKs/WalletWasabi}} that enables participation in CoinJoin rounds, the outcome of which is a CoinJoin transaction. Wasabi is based on the ZeroLink coordination protocol (\cite{zerolink}), an implementation of Chaumian CoinJoin, which allows entrusting the coordination of a round to a coordinator without needing to trust them. In this implementation, each participant receives a single post-mix output close to 0.1 in denomination and a potential change. The coordinator is compensated with a fee output in each CoinJoin transaction they finalize. A round is characterized by a post-mix denomination $d$, a coordinator fee $f$, and a number of participants, the \textit{anonymity set}, $n$. Throughout the round, participants interact with the coordinator during various phases, culminating in a CoinJoin transaction. It is possible to deduce some properties from the resulting transaction $\Delta$. Let $n$ and $d$ respectively denote the number of participants and the post-denomination.

\begin{enumerate}
\small

\item As mentioned previously, the denomination $d$\footnote{The post-denomination of $\Delta$ is generally different from $d$ since it directly depends on the TXOs provided as input by the participants as well as an estimation of the network fees at the time of the transaction's construction. However, the TXO selection process tends to ensure that the post-denomination remains close to $d$. \label{footnote_denom}} is a value close to 0.1 bitcoin. However, this value is not fixed in the protocol and tends to decrease slightly over the rounds\footnote{This gradual decrease in denomination is intentional and allows a participant to chain CoinJoin rounds without needing to add new funds.}. Nonetheless, we can assume that almost surely,
\begin{equation}
\small
\boxed{
0.1 - \epsilon \leq d \leq 0.1 + \epsilon
}
\label{eq:wasabi_approx}
\end{equation}
where $\epsilon \ll 1$ is independent of both the round and the transaction.

\item Since $n$ users participate in this transaction, the inputs are protected by at least $n$ different scripts. Moreover, there is a limit on the maximum number of inputs per participant, denoted $a_\text{max}$:

\begin{equation}
\small
\boxed{
n \leq n_\text{scripts, in} \leq | \Delta_\text{in} | \leq a_\text{max} \times n
}
\label{eq:wasabi1_n_sources}
\end{equation}

\item Each participant receives a post-mix output as well as a change output if its value is significant. Thus, we can easily provide an upper bound on the number of outputs.

\begin{equation}
\small
| \Delta_\text{out} | \leq \underbrace{n}_{\substack{\text{Number of post-mix} \\ \text{outputs}}} + \underbrace{n}_{\substack{\text{Maximum number of} \\ \text{change outputs}}} + \underbrace{1}_\text{coordinator fee}
\end{equation}

or, equivalently,

\begin{equation}
\small
\boxed{
n \geq \frac{|\Delta_\text{out}| - 1}{2}
}
\label{eq:wasabi1_bound_1}
\end{equation}

\item To fulfill its purpose, all outputs must be protected by distinct scripts:

\begin{equation}
\small
\boxed{
| \Delta_\text{out} | = n_\text{scripts, out}
}
\label{eq:wasabi1_num_scripts_out}
\end{equation}

\end{enumerate} 

\vspace{0.5cm}

Let there be a transaction  $\Delta^\prime$. We will assume that  $\Delta^\prime$ is a CoinJoin transaction of the Wasabi 1.0 type if conditions \ref{eq:wasabi_approx}, \ref{eq:wasabi1_n_sources}, \ref{eq:wasabi1_bound_1}, and \ref{eq:wasabi1_num_scripts_out} are met. Note that $n$ and $d$ are not known a priori, so they need to be estimated. For $n$, we will again use equation \ref{eq:join_estimation_n}. The observant reader will have noticed that we are overestimating $n$ in the case, although improbable but theoretically possible, where all change outputs as well as the coordinator fee output have the same values. Fortunately, the implementation implies that at least one of the participants will not have any change\footnote{The final denomination (see footnote \ref{footnote_denom}) is defined as the minimum value of the input value minus network fees provided by a participant. This directly implies that this participant will not have any change.}, so equation \ref{eq:join_estimation_n} remains valid. As for the denomination $d$, we know that it verifies

\begin{equation}
\small
d \in \left\{ v^\prime \in \mathbb{N} \ | \ \sum_{(\ket{p}, v) \in \Delta_\text{out}} 1_{v = v^\prime} = n \right\} \overset{Not.}{=} \mathcal{D}
\end{equation}

We will estimate it by taking the element from this set closest to 0.1, i.e.

\begin{equation}
\small
\hat{d} = \arg \underset{d^\prime \in \mathcal{D}}{\text{min}} \left| d^\prime - 0.1 \right| 
\label{eq:wasabi_estimate_d}
\end{equation}

\subsection{Wasabi 1.1}
\label{sec:wasabi11}
Wasabi 1.1 is software implemented in version 1.1.0 of the Wasabi wallet. This new version of Wasabi operates in a manner largely similar to the previous version but introduces mixing levels. Each mixing level represents a potential denomination for the post-mix outputs. The first mixing level, or base level, represents the basic denomination $d$, roughly equal to 0.1, already available in version 1.0. The subsequent levels are increasing multiples of this denomination; in particular, level $i$ represents the denomination $2^i \times d$. The introduction of these levels allows users to mix a larger quantity in a single transaction. Indeed, each participant can receive, for each mixing level, a post-mix output with the denomination of that level. More specifically, a user can receive an output for each level starting from the base level up to a level $j$ if their funds allow, i.e., if their funds exceed the sum of the denominations $d \times \sum_{i=0}^j 2^i$ plus the network fees. In order to maintain ambiguity for each output denomination, a level can only be used if at least two participants receive outputs of that level. A round is characterized by a denomination $d$, a coordinator fee $f$, an anonymity set $n$, and a maximum level $L$. $L$ defines the maximum level that can be used. The accessible post-mix denominations are therefore $d, 2 \times d, \cdots, 2^L \times d$\footnote{This is in fact approximately true; the final denomination of the base level depends, as in version 1.0, on the funds of each participant as well as network fees, but the denominations of the other levels remain identical to those defined by $d$.}. It is possible to deduce some properties from the resulting transaction $\Delta$. We will denote $n$ and $d$ as the number of participants and the post-denomination of the base level, respectively.

\begin{enumerate}
\small

\item As in version 1.0, the denomination of the base level is almost certainly close to 0.1, so we will assume that condition \ref{eq:wasabi_approx} remains valid here. From this, we deduce that the denomination of level $i$ satisfies
\begin{equation}
\underbrace{2^i \times (0.1 - \epsilon)}_{\overset{Not.}{=} \alpha_i} \leq 2^i \times d \leq \underbrace{2^i \times (0.1 + \epsilon)}_{\overset{Not.}{=} \beta_i}
\end{equation}
We can bound the number of outputs of level $i$ by 

\begin{equation}
\small
 \sum_{(\ket{p}, v) \in \Delta_\text{out} } 1_{\alpha_i \leq  v \leq \beta_i}
\end{equation}

which allows us to bound the number of post-mix outputs by

\begin{equation}
\small
\sum_{i=0}^{L} \left( \sum_{(\ket{p}, v) \in \Delta_\text{out}} 1_{\alpha_i \leq v \leq \beta_i} \right)
\end{equation}

We can then bound the number of outputs

\begin{equation}
\small
| \Delta_\text{out} | \leq \underbrace{\sum_{i=0}^{L} \left( \sum_{(\ket{p}, v) \in \Delta_\text{out}} 1_{\alpha_i \leq v \leq \beta_i} \right) }_{\substack{\text{Maximum number of} \\ \text{post-mix outputs}}} + \underbrace{n}_{\substack{\text{Maximum number of} \\ \text{change outputs}}} + 1
\end{equation}

or, equivalently

\begin{equation}
\small
\boxed{
\sum_{i=0}^{L} \left( \sum_{(\ket{p}, v) \in \Delta_\text{out}} 1_{\alpha_i \leq v \leq \beta_i} \right)  \geq | \Delta_\text{out} |  -  n - 1
}
\label{eq:wasabi11_bound}
\end{equation}

\item Conditions \ref{eq:wasabi1_n_sources} and \ref{eq:wasabi1_num_scripts_out} remain valid.

\end{enumerate}

Let there be a transaction $\Delta^\prime$, we will assume that $\Delta^\prime$ is a CoinJoin transaction of the Wasabi 1.1 type if conditions \ref{eq:wasabi_approx}, \ref{eq:wasabi1_n_sources}, \ref{eq:wasabi1_num_scripts_out}, and \ref{eq:wasabi11_bound} are met. Again, $n$ and $d$ are not known a priori. As with Wasabi 1.0, at least one participant will not receive change. Moreover, each participant receives an output from the base level, and at most one output from the following levels, so equation \ref{eq:join_estimation_n} remains valid for estimating $n$. We will then use equation \ref{eq:wasabi_estimate_d} to estimate $d$.

\subsection{Wasabi 2.0}

\label{sec:wasabi2}

Wasabi 2.0, implemented in version 2.0.0.0 of the Wasabi Wallet, is the result of a major software update. This update introduces fixed denominations and the use of the WabiSabi coordination protocol (\cite{ficsor2021wabisabi}). WabiSabi is an implementation of a generalized Chaumian CoinJoin, replacing the ZeroLink protocol from versions 1.0 and 1.1. A CoinJoin round is characterized by a set of denominations $\mathcal{D}$\footnote{In reality, these denominations are not consistently all usable; the usable denominations are determined during the round by the participants' inputs $\Delta_\text{in}$.}, a coordinator fee $f$, and a target number of inputs $p$. Each participant receives as output a combination of post-mix outputs with denominations from $\mathcal{D}$, as well as a potential change. Unlike previous versions, a denomination can be chosen multiple times by the same participant. Wasabi 2.0 also implements an optimization algorithm to select the best decomposition from all possible decompositions of the input value into denominations of $\mathcal{D}$. The optimization criteria are numerous, particularly prioritizing the absence of change. It is possible to deduce some properties from the resulting transaction $\Delta$. We will denote $n$, $n_\text{post-mix}$ and $n_\text{change}$ as the number of participants, the number of post-mix outputs, and the number of changes, respectively.
\begin{enumerate}
\small

\item A participant receives at least one post-mix output and at most one change, from which we deduce that

\begin{equation}
\small
n_\text{change} \leq n \leq n_\text{post-mix}
\end{equation} 

Each mix-output has a denomination belonging to $\mathcal{D}$, thus

\begin{equation}
\small
n_\text{post-mix} \leq \sum_{(\ket{p}, v) \in \Delta_\text{out}} 1_{v \in \mathcal{D}}
\end{equation}

Conversely, any output that doesn't have a standardized value can only be a change output or the coordinator fee, thus

\begin{equation}
\small
n_\text{change} \geq \sum_{(\ket{p}, v) \in \Delta_\text{out}} 1_{v \notin \mathcal{D}} - 1
\end{equation}

From the three previous inequalities, we deduce that

\begin{equation}
\sum_{(\ket{p}, v) \in \Delta_\text{out}} 1_{v \notin \mathcal{D}} - 1 \leq \sum_{(\ket{p}, v) \in \Delta_\text{out}} 1_{v \in \mathcal{D}}
\end{equation}

Or, 

\begin{equation}
\small
\sum_{(\ket{p}, v) \in \Delta_\text{out}} 1_{v \in \mathcal{D}} + \sum_{(\ket{p}, v) \in \Delta_\text{out}} 1_{v \notin \mathcal{D}} = | \Delta_\text{out} | 
\end{equation}

Thus

\begin{equation}
\small
\boxed{
\sum_{(\ket{p}, v) \in \Delta_\text{out}} 1_{v \in \mathcal{D}} \geq \frac{| \Delta_\text{out} | - 1 }{2}
}
\label{eq:wasabi2_bound}
\end{equation}

\item By definition of the target number of inputs,

\begin{equation}
\small
\boxed{
| \Delta_\text{in} | \geq p
}
\label{eq:wasabi2_bound_target}
\end{equation}

\item Wasabi 2.0 has introduced a limit $a_\text{max}$\footnote{In fact, this is a standard configuration parameter, however, nothing prevents the user from changing it.}

\begin{equation}
\small
n \geq \frac{| \Delta_\text{in} | }{a_\text{max}} 
\end{equation}

Thus 

\begin{equation}
\small
\boxed{
\sum_{(\ket{p}, v) \in \Delta_\text{out}} 1_{v \in \mathcal{D}}  \geq \frac{| \Delta_\text{in} | }{a_\text{max}} 
}
\label{eq:wasabi2_bound_outputs}
\end{equation}

\item Furthermore, the protocol has implemented a minimum value $v_\text{min}$ for each output

\begin{equation}
\small
\boxed{
\underset{(\ket{p}, v) \in \Delta_\text{in}}{\text{min}} v \geq v_\text{min}
}
\label{eq:wasabi2_min_denom_input}
\end{equation}

\item Finally, for the CoinJoin to fulfill its purpose, all output scripts must be distinct:

\begin{equation}
\small
\boxed{
n_\text{scripts, out} = | \Delta_\text{out} | 
}
\label{eq:wasabi2_script_out}
\end{equation}

\end{enumerate}

\vspace{0.5cm}

Let there be a transaction  $\Delta^\prime$, we will assume that  $\Delta^\prime$ is a Wasabi 2.0 type CoinJoin transaction if conditions \ref{eq:wasabi2_bound}, \ref{eq:wasabi2_bound_target}, \ref{eq:wasabi2_bound_outputs}, \ref{eq:wasabi2_min_denom_input}, and \ref{eq:wasabi2_script_out} are satisfied.

\subsection{Whirlpool} 

\label{sec:whirlpool}

 Whirlpool\footnote{\url{https://code.samourai.io/whirlpool}} is a software implemented in the Samourai\footnote{\url{https://code.samourai.io/wallet/samourai-wallet}} and Sparrow \footnote{\url{https://sparrowwallet.com/}} wallets that allows their users to participate in CoinJoin rounds. Whirlpool also entrusts the coordination of rounds to a coordinator using a Chaumian CoinJoin implementation forked from ZeroLink. CoinJoin rounds are organized by pool, with each pool characterized by a denomination $d$ and a coordinator fee $f$. The four currently available pools can be seen in Table \ref{table:samourai_pool}

 \begin{table}[h!]
\begin{center}
\begin{tabular}{|c|c|}
\hline
Denomination  & Protocol Fee  \\
\hline
0.001 & 0.00005 \\
0.01 & 0.0005 \\
0.05 & 0.00175\\
0.5 & 0.0175 \\
\hline
\end{tabular}
\end{center}
\caption{Samourai pools. Quantities are expressed in bitcoins.}
\label{table:samourai_pool}
\end{table}%

In order to participate in CoinJoin transactions within a pool $(d, f)$, a user must first convert funds in their possession into outputs of value $d + \epsilon > d$ during a Tx0-type transaction. During this transaction, the participant receives $n_0$ outputs of value $d + \epsilon$ and a potential change representing the value that could not be converted. The coordinator only compensates during this type of transaction by adding a coordinator fee output of value $f$. Since the outputs of values $d + \epsilon$ have not been mixed at this stage, we will refer to them as pre-mix. The surplus $\epsilon$ compared to the pool denomination will be used to pay the network fee during the first mix. For this reason, $\epsilon$ is not fixed and depends on an estimate of the network fees at the time the transaction is constructed.\\
 
 We will denote $\mathcal{P}$ as the set of pools. Let $(d, f) \in \mathcal{P}$ a pool and  $\Delta$ a  Tx0 transaction for this pool.
 
 \begin{enumerate}
 \small
 
 \item The output set of $\Delta$  consists of the post-mix outputs of value $d + \epsilon$, an output of value $f$ to compensate the coordinator, a zero-value output used to store data, and possibly one last output to receive the change that could not be converted. From this, we deduce that at most 3 outputs are not pre-mix outputs, thus
 
 \begin{equation}
 \small
 \boxed{
 \sum_{(\ket{p}, v) \in \Delta_\text{out}} 1_{v = d + \epsilon} \geq | \Delta_\text{out}| - 3
 }
 \label{eq:whirlpool_bound_non_pre_mix}
 \end{equation}
 
 \item $\Delta_\text{out}$ consists of at least one pre-mix output, one coordinator fee output, and one zero-denomination output, therefore
 
 \begin{align}
 \small
 \sum_{(\ket{p}, v) \in \Delta_\text{out}} 1_{v = d + \epsilon} \geq 1 \label{eq:samourai_exists_premix} \\
 \sum_{(\ket{p}, v) \in \Delta_\text{out}} 1_{v = f} = 1 \label{eq:samourai_exists_fee} \\ 
  \sum_{(\ket{p}, v) \in \Delta_\text{out}} 1_{v = 0} = 1 \label{eq:samourai_exists_opreturn}
\end{align}

However, Samourai has implemented a coupon system to pay lower fees during Tx0. In addition, the fees may have changed over time, so we relax condition \ref{eq:samourai_exists_fee} to

 \begin{equation}
 \small
 \boxed{
\sum_{(\ket{p}, v) \in \Delta_\text{out}} 1_{\eta_1 \times f \leq v \leq  \eta_2 \times f} = 1 \label{eq:samourai_exists_fee_relaxed}
}
\end{equation}

where $\eta_1 \in [0, 1]$ and $\eta_2 \geq 1$ . 

\item The protocol has implemented a limit $a_\text{max}$ on the maximum number of pre-mix outputs:

 \begin{equation}
 \small
 \boxed{
 \sum_{(\ket{p}, v) \in \Delta_\text{out}} 1_{v = d + \epsilon} \leq a_\text{max}
 }
 \label{eq:samourai_bound_premix}
\end{equation}

\item $\epsilon$ is not fixed and depends on an estimate of network fees at the time of the transaction. However, Whirlpool has implemented a range of acceptable values for $\epsilon$,

\begin{equation}
\small
\boxed{
\epsilon_\text{min} \leq \epsilon \leq \epsilon_\text{max}
}
\label{eq:samourai_bound_epsilon}
\end{equation}
 
 \end{enumerate} 
 
 \vspace{0.5cm}
 
Let there be a transaction  $\Delta^\prime$, we will assume that $\Delta^\prime$ is a Tx0 transaction if conditions \ref{eq:whirlpool_bound_non_pre_mix}, \ref{eq:samourai_exists_premix}, \ref{eq:samourai_exists_opreturn}, \ref{eq:samourai_exists_fee_relaxed}, \ref{eq:samourai_bound_premix}, and \ref{eq:samourai_bound_epsilon} are satisfied. Here, we need to estimate $d$, $\epsilon$, and $f$. According to \ref{eq:samourai_bound_epsilon}, an output with a value of $v$ can be a pre-mix denomination if
 
 \begin{equation}
 \small
 \exists (d, f) \in \mathcal{P}, \ v \in [d + \epsilon_\text{min}, d + \epsilon_\text{max}]
 \end{equation}
 
Among the output values in $\Delta$ that satisfy the previous condition, we will estimate the pre-mix denomination as the output value with the most occurrences in $\Delta_\text{out}$. In case of a tie, we will take the highest value. Let $\tilde{d}$ be the selected value; then,

 \begin{equation}
 \small
 (\hat{d}, \hat{f}) = \arg \underset{(d, f) \in \mathcal{P}, \ d \leq \tilde{d}}{\text{min}} | d - \tilde{d}|
 \end{equation}
 and $\tilde{\epsilon} = \tilde{d} - d$. \\

A CoinJoin or mix transaction involves exactly five distinct users. Each user participates in the round by contributing a TXO from either a previous mix or a Tx0 transaction and receives a post-mix output with the denomination of the pool. The coordination of a round is carried out in a manner similar to Wasabi.

\begin{enumerate}
\small

\item $\Delta$ has exactly 5 inputs from distinct scripts and 5 outputs destined for five distinct scripts.

\begin{equation}
\small
\boxed{
| \Delta_\text{in}| = n_\text{scripts, in} = n_\text{scripts, out} = | \Delta_\text{out}| = 5
}
\label{eq:samourai_num_scripts}
\end{equation}

\item All the post-mix outputs have a denomination from an existing pool.

\begin{equation}
\small
\boxed{
\exists (d, f) \in \mathcal{P}, \ \sum_{(\ket{p}, v) \in \Delta_\text{out}} 1_{v = d} = 5
}
\label{eq:samourai_denom_output}
\end{equation}

and all inputs are either outputs from a previous mix or outputs from Tx0, so

\begin{equation}
\small
\boxed{
\sum_{(\ket{p}, v) \in \Delta_\text{in}} 1_{v \in [d, d + \epsilon_\text{max}]} = 5
}
\label{eq:samourai_denom_input}
\end{equation}

\item In order to allow post-mix outputs to be remixed regularly, the protocol requires at least one input from a previous mix. Additionally, it is necessary to have at least one pre-mix output to cover the network fees, thus:

\begin{equation}
\small
\boxed{
1 \leq \sum_{(\ket{p}, v) \in \Delta_\text{in}} 1_{v > d} \leq 4
}
\label{eq:samourai_bound_premix2}
\end{equation}

\end{enumerate}

Let a transaction be $\Delta^\prime$, we will assume that  $\Delta^\prime$ is a Whirlpool CoinJoin transaction if conditions \ref{eq:samourai_num_scripts}, \ref{eq:samourai_denom_output}, \ref{eq:samourai_denom_input},  and \ref{eq:samourai_bound_premix2}  are satisfied.

\section{Detection Results}

In the previous section, we discussed the properties of transactions originating from JoinMarket, Wasabi 1.0, Wasabi 1.1, Wasabi 2.0, and Whirlpool, based on the implementation of each protocol. We also detailed how one can determine from a transaction $\Delta$ if it comes from any of the studied protocols. Since the conditions we developed aim to be necessary conditions, we can expect a low rate of false negatives. Conversely, the rate of false positives might be significant if other CoinJoin protocols not studied here have similar transaction structures. However, we endeavored to derive precise conditions to minimize false positives. In this section, we count the number of transactions meeting the criteria from block 0 to block 760,000. To extract data from the Bitcoin blockchain, we set up a Bitcoin Core full node. This required downloading and syncing the complete transaction ledger from a network of peers. After installing the latest Bitcoin Core software and configuring the node, the entire transaction history was saved in the local blockchain data directory, specifically in the "blkXXXXX.dat" files in the "/.blocks" folder. We then delved into the Bitcoin protocol and file structure, using parsing techniques to extract transaction details. This process ensured accurate data for our analysis.

\paragraph{Number of JoinMarket transactions} 

\begin{figure}[H]
    \centering
    \includegraphics[width=0.75\textwidth]{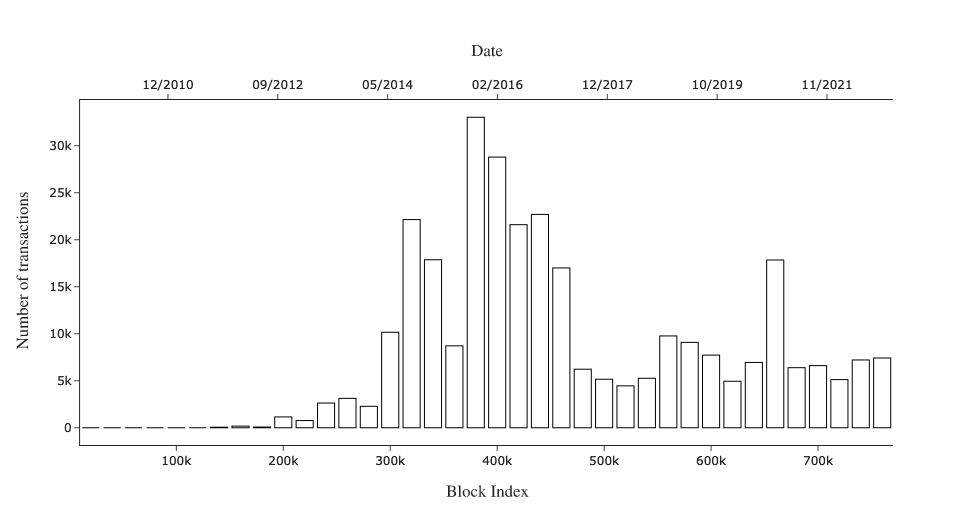}
    \caption{For different block indices $k$ ($x$-axis), number of transactions detected ($y$-axis) by the JoinMarket heuristic between the blocks with indices $k - 20,000$ and $k$. The different indices considered are spaced by 20,000.}
    \label{fig:joinmarket_last}
\end{figure}

\begin{figure}[H]
    \centering
    \includegraphics[width=0.75\textwidth]{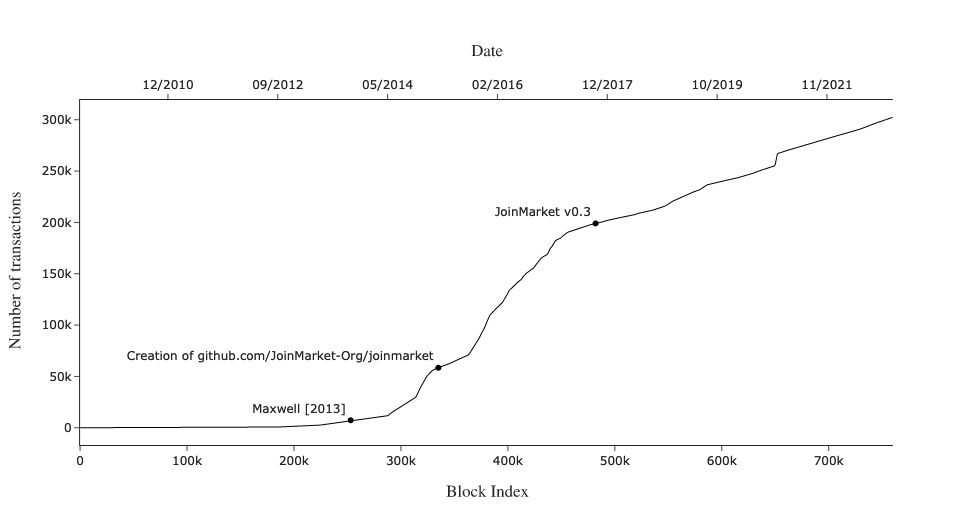}
    \caption{Cumulative number of transactions ($y$-axis) detected by the JoinMarket heuristic as a function of the last block index considered ($x$-axis).}
    \label{fig:joinmarket_cumsum}
\end{figure}

In this section, we count the transactions that meet the criteria of a JoinMarket transaction (see section \ref{sec:joinmarket}). Since transactions of type Wasabi 1.0 or Whirlpool can also meet these criteria, we will not count them in this section. Furthermore, as the format identified at the end of section \ref{sec:joinmarket} is relatively broad – a standardized output and a change output for each participant – it is likely we will detect transactions from other protocols or softwares. The quantities found can then be considered a proxy for CoinJoin transactions not studied in our research. To measure the evolution of JoinMarket's popularity over time, we counted, for different block indices, the number of transactions detected between the blocks with indices $k - 20,000$ and $k$. We plotted the result obtained for different values of $k$ in Figure  \ref{fig:joinmarket_last}. We also plotted in Figure \ref{fig:joinmarket_cumsum} the cumulative number of detected JoinMarket transactions as a function of the last  block index considered. The significant number of transactions detected before the GitHub repository of the project was created suggests that our heuristic indeed captures transactions from other protocols. In particular, we observe a surge in the number of CoinJoin transactions following Gregory Maxwell's post on the BitcoinTalk forum. This type of CoinJoin seems to have then lost in popularity, perhaps due to the arrival of other CoinJoin implementations.

\paragraph{Number of Wasabi 1 transactions} 

\begin{figure}[H]
    \centering
    \includegraphics[width=0.75\textwidth]{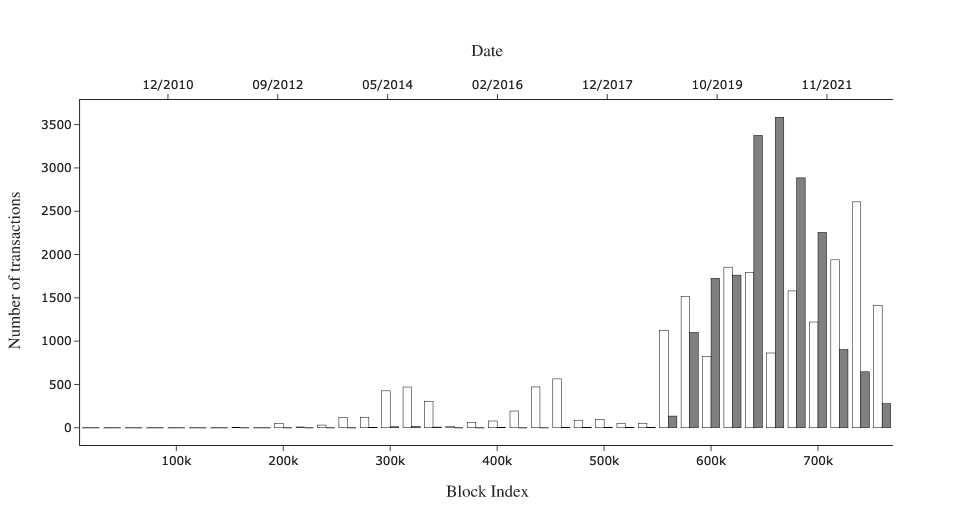}
    \caption{For different block indices $k$ ($x$-axis), number of transactions detected ($y$-axis) by the Wasabi 1.0/1.1 heuristics between the blocks with indices $k - 20,000$ and $k$. The different indices considered are spaced by 20,000. White bars corresponds to the number of single-denomination transactions, and grey bars to the number of multi-denomination transactions.}
    \label{fig:wasabi1_last}
\end{figure}

\begin{figure}[H]
    \centering
    \includegraphics[width=0.75\textwidth]{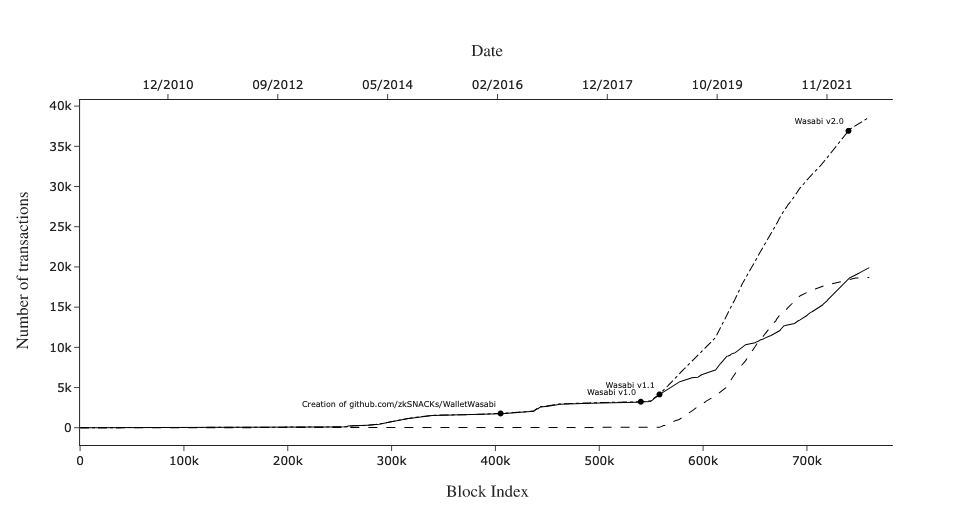}
    \caption{Cumulative number of transactions ($y$-axis) detected by the Wasabi 1.0/1.1 heuristics as a function of the block index ($x$-axis): single-denomination (solid line), multi-denomination (dashed line), total (dot-dashed line).}
    \label{fig:wasabi1_cumsum}
\end{figure}

In this section, we count the transactions that satisfy the criteria for a Wasabi 1.0 transaction (see section \ref{sec:wasabi1}) or a Wasabi 1.1 transaction (see section \ref{sec:wasabi11}). It is worth noting that a Wasabi 1.0 transaction also meets the criteria for a Wasabi 1.1 transaction. For this reason, in this section, we will rather distinguish between single-denomination transactions and multi-denomination transactions (introduced with Wasabi 1.1). We have plotted in Figure \ref{fig:wasabi1_last} for different values of $k$ the number of single-denomination and multi-denomination transactions detected between the blocks with indices $k - 20,000$ and $k$. In Figure \ref{fig:wasabi1_cumsum}, we charted the cumulative number of detected Wasabi transactions as a function of the last block index considered. We observed a significant number of single denomination transactions detected before the creation of the project's GitHub repository. This is not surprising as the Wasabi 1.0 transactions resemble a JoinMarket transaction with a denomination close to 0.1, we may then detect some transactions that we have excluded in the previous section. Conversely, the spike in multi-denomination transactions aligns perfectly with the release of version 1.1, and the decline of multi-denomination transactions aligns with the release of version 2.0, which is highly consistent.

\paragraph{Number of Wasabi 2 transactions}  

\begin{figure}[H]
    \centering
    \includegraphics[width=0.75\textwidth]{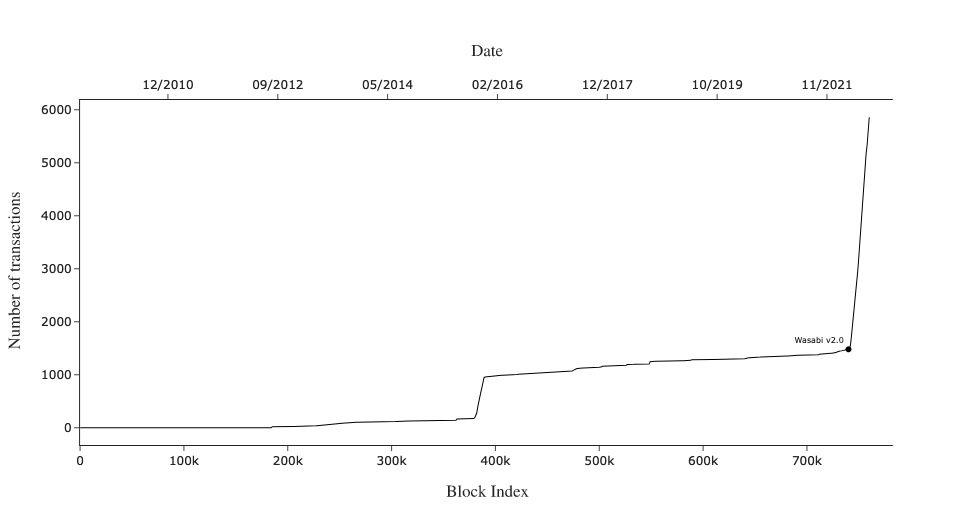}
    \caption{Cumulative number of transactions ($y$-axis) detected by the Wasabi 2.0 heuristic as a function of the block index ($x$-axis).}
    \label{fig:wasabi2_cumsum}
\end{figure}

In this section, we count the transactions that meet the criteria for a Wasabi 2.0 transaction. We will use the parameters  $a_\text{max}=10$, $p=50$, and $v_\text{min}=5000$. In Figure \ref{fig:wasabi2_cumsum}, we plotted the cumulative number of Wasabi 2.0 transactions detected based on the criteria established in section \ref{sec:wasabi2} as a function of the block index. It is worth noting a number of false positives before version 2.0, particularly between blocks 375,000 and 400,000. However, the surge in detected transactions aligns well with the release of the 2.0 version of the protocol.

\paragraph{Number of  Whirlpool transactions}

\begin{figure}[H]
    \centering
    \includegraphics[width=0.75\textwidth]{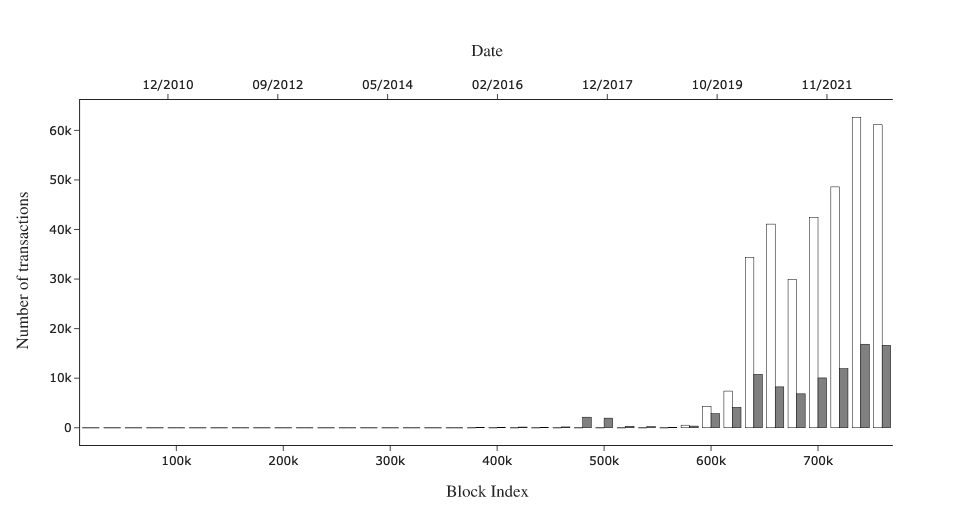}
    \caption{For different block indices $k$ ($x$-axis), number of transactions detected ($y$-axis) by the Whirlpool heuristics between the blocks with indices $k - 20,000$ and $k$. The different indices considered are spaced by 20,000. White bars corresponds to the number of Tx0 transactions, and grey bars to the number of CoinJoin transactions.}
    \label{fig:samourai_last}
\end{figure}

\begin{figure}[H]
    \centering
    \includegraphics[width=0.75\textwidth]{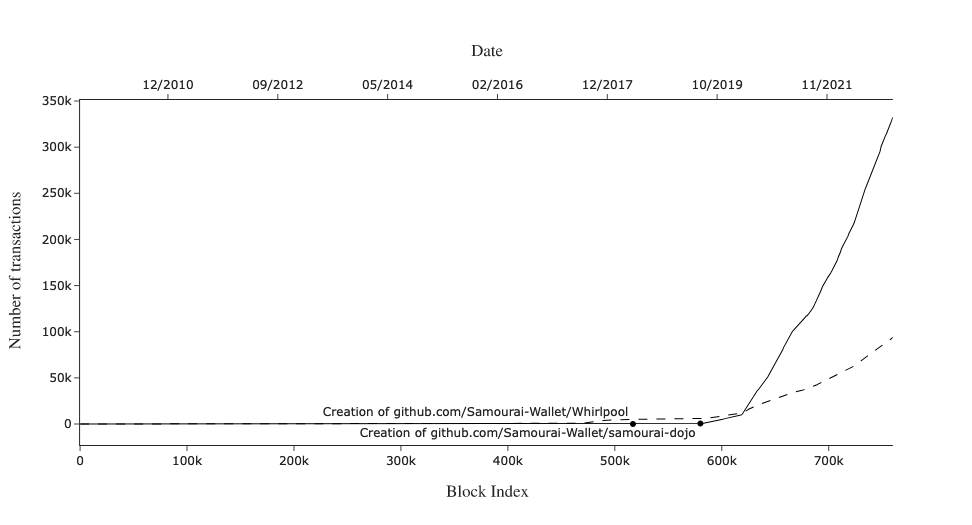}
    \caption{Cumulative number of transactions ($y$-axis) detected by the Whirlpool heuristics as a function of the block index ($x$-axis): Tx0 (solid line), CoinJoin (dashed line).}
    \label{fig:samourai_cumsum}
\end{figure}

In this section, we count the transactions that meet the criteria for a Tx0 or CoinJoin transaction of the Whirlpool protocol. For this purpose, we employed the conditions from the section with the parameters $a_\text{max}=70$, $\eta_1=0.5$, $\eta_2=3$, $\epsilon_\text{min}=100$ and $\epsilon_\text{max}=100000$. We have plotted in Figure \ref{fig:samourai_last} for different values of $k$ the number of transactions of both types detected between the blocks with indices $k - 20,000$ and $k$. In Figure \ref{fig:samourai_cumsum}, we charted the cumulative number of detected CoinJoin transactions for both types as a function of the last block index considered. We can observe that the number of both types of transactions surged simultaneously starting from block 560,000, which aligns with the initial releases of the various GitHub repositories of the project.

\begin{figure}[H]
    \centering
    \includegraphics[width=0.75\textwidth]{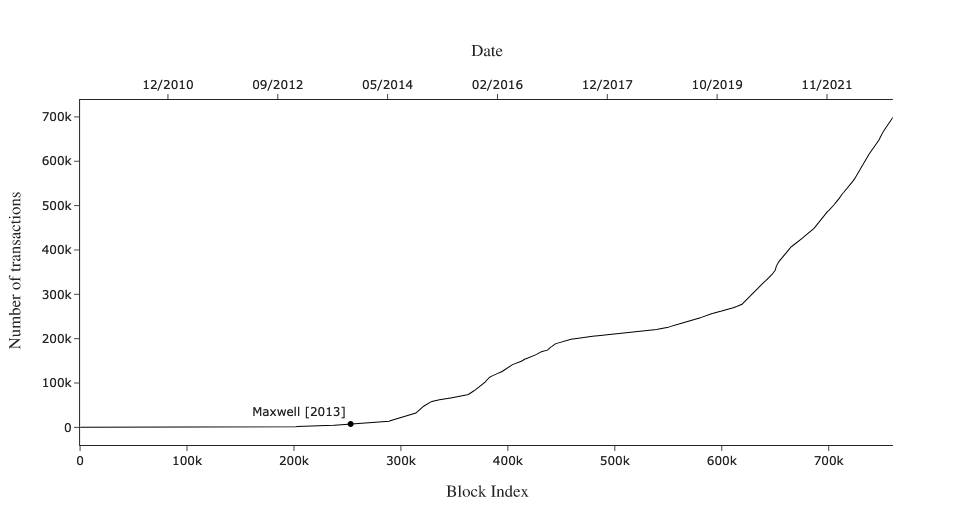}
    \caption{Cumulative number of transactions ($y$-axis) detected by at least one heuristic as a function of the block index ($x$-axis).}
    \label{fig:total_cumsum}
\end{figure}

\paragraph{Number of CoinJoin transactions} Finally, we count the number of transactions that meet the criteria of at least one heuristic, excluding Whirlpool Tx0 transactions. We have plotted in Figure \ref{fig:total_cumsum} the cumulative number of detected CoinJoin transactions as a function of the last block index considered.

\label{sec:results}

\section*{Conclusion}

Throughout this study, we meticulously developed a series of heuristics aimed at detecting transactions stemming from protocols that facilitate CoinJoins. Drawing directly from the open-source implementations of these protocols and software tools provided the foundation for our heuristic approach. The intricacy of this endeavor lies in the absence of a comprehensive dataset of labeled CoinJoin transactions, making validation a non-trivial task. Nonetheless, our findings in section 3 exhibit a significant degree of consistency, even when accounting for the inevitable presence of false positives. Analyzing the data, we discerned an upward trajectory in the overall volume of CoinJoin transactions. This growth not only reflects the maturation and robustness of the network over time but also underscores a mounting sentiment among participants for enhanced privacy in their transactions. This underlines the critical importance and relevance of our research, given the evolving demands and complexities of the Bitcoin network.

\bibliography{biblio}

\end{document}